\documentclass{article}
\usepackage{amsfonts}
\usepackage{epic}
\def\cA{{\cal A}}          \def\cB{{\cal B}}          
          \def\cE{{\cal E}}          \def\cF{{\cal F}}
                    
          \def\cK{{\cal K}}          \def\cL{{\cal L}} 
\def\cM{{\cal M}}                    
                     
                    \def\cU{{\cal U}}
          \def\cW{{\cal W}}

                    \def\bL{{\bar L}}

\def\CC{{\mathbb C}}

\def\ZZ{{\mathbb Z}}
\newcommand{\un}{\mbox{1\hspace{-1mm}I}}
\newcommand{\si}{\sigma}

\newcommand{\nn}{\nonumber}

\begin{document}

\pagestyle{empty}

\begin{center} 
\textsf{\Large 
{Generalization of the $\cU_q(gl(N))$ algebra and staggered models}} 
  
\vspace{36pt}
{\bf D.~Arnaudon\footnote{e-mail:{\sl arnaudon@lapp.in2p3.fr}}},
{\bf A.~Sedrakyan\footnote{e-mail:{\sl sedrak@lx2.yerphi.am,
Permanent address: Yerevan Physics Institute, Armenia}}},
{\bf T.~Sedrakyan\footnote{e-mail:{\sl tigrans@moon.yerphi.am,
Permanent address: Yerevan Physics Institute, Armenia}}},
{\bf P.~Sorba\footnote{e-mail:{\sl sorba@lapp.in2p3.fr}}}\\  

\vspace{30pt}

\emph{Laboratoire d'Annecy-le-Vieux de Physique Th{\'e}orique LAPTH}
\\
\emph{CNRS, UMR 5108, associ{\'e}e {\`a} l'Universit{\'e} de Savoie}
\\
\emph{BP 110, F-74941 Annecy-le-Vieux Cedex, France}
\\

\vfill
{\bf Abstract}
\end{center}

We develop a technique of construction of integrable models
with a $\ZZ_2$ grading of both the auxiliary (chain) and quantum (time)
spaces. These models have a staggered disposition of the anisotropy
parameter. The corresponding Yang--Baxter Equations are written
down and their solution for the $gl(N)$ case are found. We analyze
in details the $N=2$ case and find the corresponding
quantum group behind this solution. It can be regarded as
quantum $\cU_{q\cB}(gl(2))$ group with a matrix deformation parameter
$q\cB$ with $(q\cB)^2=q^2$.
The symmetry behind these models can also be interpreted as the tensor
product of the (-1)-Weyl algebra by an extension of $\cU_q(gl(N))$
with a Cartan generator related to deformation parameter~-1. 

\vfill
\rightline{LAPTH-855/01}
\rightline{hep-th/0106139}
\rightline{June 2001}

\newpage
\pagestyle{plain}
\setcounter{page}{1}

\section{Introduction}
\setcounter{equation}{0}

\indent

In some physical problems \cite{S1,CC} we have to work with a 
partition function defined by the action on the Manhattan Lattice
$(ML)$. The geometry of $(ML)$ automatically defines a chess like 
structure for the action. Therefore, when one defines the 
quantum chain Hamiltonian from this type of action by use of coherent
states \cite{F}, as it was done in the article \cite{S1}, we 
necessarily come to the idea of $\ZZ_2$ grading of the spaces
along the chain and time directions. Precisely we have 
alternating auxiliary spaces in the chain direction  and 
alternating quantum spaces
in the time directions, defined as in article \cite{APSS}

The aim of the present article is the definition of the
corresponding $\ZZ_2$-structure, the formulation of the conditions
(Yang--Baxter Equations $(YBE)$) under which we can construct
an integrable model. We have found that the corresponding 
extended 
YBE's has a solution
for general $gl(N)$ group and that integrable models
with  staggered parameter anisotropy can be constructed.
We study in details the quantum group structure
behind this construction for the case of $\cU_q(gl(2))$. 
For spin-1/2 representation of $gl(2)$ group (XXZ chain)
this type of model 
was defined and studied in the article \cite{APSS} and for
spin-1 degrees of freedom (anisotropic $t-J$
model) in \cite{ASSS}.

\section{Basic Definitions}
\setcounter{equation}{0}

\indent

Let us now consider $\ZZ_2$ graded quantum $V_{j,\rho}(v)$ 
(with $j=1,.....N$ as a chain index) and 
auxiliary $V_{a,\si}(u)$ spaces, where $\rho, \si =0,1$ are
the grading indices. Consider
$R$-matrices, which act on the direct product
of  spaces $V_{a,\si}(u)$ and $ V_{j,\rho}(v)$, $(\si,\rho =0,1)$,
mapping them on the intertwined direct product of 
$V_{a,\bar{\si}}(u)$ and $ V_{j,\bar{\rho}(v)}$ with the complementary
$\bar{\si}=(1-\si)$, $\bar{\rho}=(1-\rho)$ indices
\begin{equation}
\label{R1}
R_{aj,\si \rho}\left( u,v\right):\quad V_{a,\si}(u)\otimes 
V_{j,\rho}(v)\rightarrow V_{j,\bar{\rho}}(v)\otimes V_{a,\bar{\si}}(u).  
\end{equation}
\medskip

\noindent
{\large \sc Definition.} 
\textsl{It is convenient to introduce 
two transmutation operations $\iota_1$
and $\iota_2$ with the property $\iota_1^2=\iota_2^2=id$ 
for the quantum and auxiliary spaces
correspondingly, and to mark the operators $R_{aj,\si\rho}$ as 
follows
\begin{eqnarray}
\label{R2}
R_{aj,00}&\equiv& R_{aj},\qquad R_{aj,01}\equiv R_{aj}^{\iota_1},\nn\\
R_{aj,10}&\equiv& R_{aj}^{\iota_2},\qquad R_{aj,11}\equiv R_{aj}^{\iota_1 
\iota_2}.
\end{eqnarray}
}

The introduction of the $\ZZ_2$ grading of quantum spaces 
in time direction means, 
that we have now two monodromy operators $T_{\rho}, \rho=0,1$,
which act on the space $V_{\rho}(u)=\prod_{j=1}^N V_{j,\rho}(u)$
by mapping it on $V_{\bar{\rho}}(u)=\prod_{j=1}^N V_{j,\bar{\rho}}(u)$
\begin{equation}
\label{T}
T_\rho(v,u) \qquad : V_\rho(u) \rightarrow V_{\bar{\rho}}(u), \qquad \qquad 
\rho=0,1.
\end{equation}

It is clear now, that the monodromy operator of the model, which is defined
by translational invariance in two steps in the time direction and  
determines the partition function, is the product of two monodromy operators
\begin{equation}
\label{TT}
T(v,u) = T_0(v,u) T_1(v,u).
\end{equation}

The $\ZZ_2$ grading of auxiliary spaces along the chain direction means
that the $T_0(u,v)$ and $T_1(u,v)$ monodromy matrices are defined
according to the following
\medskip

\noindent
{\large \sc Definition.} 
\textsl{We define the monodromy operators $T_{0,1}(v,u)$ 
as a staggered product
of the $R_{aj}(v,u)$ and $\bar{R}_{aj}^{\iota_2}(v,u)$ matrices:
\begin{eqnarray}
\label{T1}
T_1(v,u)=\prod_{j=1}^N R_{a,2j-1}(v,u)
\bar{R}_{a,2j}^{\iota_2}(v,u)\nn\\
T_0(v,u)=\prod_{j=1}^N \bar{R}_{a,2j-1}^{\iota_1}(v,u)
R_{a,2j}^{\iota_1 \iota_2}(v,u),
\end{eqnarray}
}
where the notation $\bar{R}$ denotes a 
different parametrization of the $R(v,u)$-matrix via spectral
parameters
$v$ and $u$ and can be considered as an operation
over $R$ with property $\bar{\bar{R}}= R$.
For the integrable models where the intertwiner matrix $R(v-u)$
simply depends
on the difference of the spectral parameters $v$ and $u$ 
this operation means the shift of its argument $u$ as follows
\begin{equation}
\label{RR}
\bar{R}(u)=R(\bar u), \qquad \bar{u}=\theta-u,
\end{equation}
where $\theta$ is an additional model parameter. We will consider
this case in this paper.

\section{Staggered Yang--Baxter equations}
\setcounter{equation}{0}

\indent

As it is well known in Bethe Ansatz Technique, the sufficient
condition for the commutativity of transfer matrices $\tau(u)=
Tr T(u)$ with different spectral parameters is the YBE. For our
case we have a two sets of equations \cite{APSS}

\begin{eqnarray}
  \label{eq:YBE1}
  R_{12}(u,v) \bar{R}_{13}^{\iota_1}(u) R_{23}(v)=
  R_{23}^{\iota_1}(v) \bar{R}_{13}(u) \tilde{R}_{12}(u,v)
\end{eqnarray}
and
\begin{eqnarray}
  \label{eq:YBE2}
  \tilde{R}_{12}(u,v) R_{13}^{\iota_1 \iota_2}(u) 
  \bar{R}_{23}^{\iota_2}(v)=
  \bar{R}_{23}^{\iota_1 \iota_2}(v) R_{13}^{\iota_2}(u) R_{12}(u,v) \;,
\end{eqnarray}
with $\bar{R}(u)\equiv R(\bar{u})$ and $R^{\iota_2}(u)=R^{\iota_1}(-u)$.

From $R(u)$ above, we  follow a procedure which is the inverse of
the Baxterisation (debaxterisation) \cite{Jones}.  
Let 
\begin{equation}
  \label{eq:deBaxterise}
  R_{12}(u) = \frac{1}{2i} \left(z R_{12} - z^{-1} R_{21}^{-1} \right)
\end{equation}
with $z=e^{iu}$ and the constant $R_{12}$ and $R_{21}^{-1}$ matrices are
spectral parameter independent. Then the Yang--Baxter equations 
(\ref{eq:YBE1})--(\ref{eq:YBE2})
for the spectral parameter dependent $R$-matrix $R(u)$ and
$R^{\iota_1}(u)$ are
equivalent to the following equations for the constant $R$-matrices

\begin{eqnarray}
  \label{eq:YBEconst1}
  R_{12} R_{13}^{\iota_1} R_{23} &=&
  R_{23}^{\iota_1} R_{13} R_{12}^{\iota_1} \\
  \label{eq:YBEconst2}
  R_{12}^{\iota_1} R_{13} R_{23}^{\iota_1} &=&
  R_{23} R_{13}^{\iota_1} R_{12} \\
  \label{eq:YBEconst3}
  R_{12} \left(R_{31}^{\iota_1}\right)^{-1} R_{23} 
  - \left(R_{21}\right)^{-1} R_{13}^{\iota_1} \left(R_{32}\right)^{-1} &=&
  R_{23}^{\iota_1} \left(R_{31}\right)^{-1} R_{12}^{\iota_1} 
  - \left(R_{32}^{\iota_1}\right)^{-1} R_{13}
  \left(R_{21}^{\iota_1}\right)^{-1}  \\
  \label{eq:YBEconst4}
  R_{12}^{\iota_1} \left(R_{31}\right)^{-1} R_{23}^{\iota_1} 
  - \left(R_{21}^{\iota_1}\right)^{-1} R_{13}
  \left(R_{32}^{\iota_1}\right)^{-1} &=& 
  R_{23} \left(R_{31}^{\iota_1}\right)^{-1} R_{12} 
  - \left(R_{32}\right)^{-1} R_{13}^{\iota_1} \left(R_{21}\right)^{-1} 
\end{eqnarray}
assuming $\tilde{R} = R^{\iota_1}$. 

If this modified YBE's have a solution, then one can formulate
a new integrable model on the basis of the existing ones. 
We will hereafter give solutions of these YBE's based on  $\cU_q(gl(N))$
$R_q$-matrices, for arbitrary~$n$.

\section{$\cU_q(gl(2))$ case\label{sect:gl2case}}
\setcounter{equation}{0}

\indent

As proved in \cite{APSS} in connection with the staggered XXZ model, a
solution of (\ref{eq:YBE1})--(\ref{eq:YBE2}) is given by 
\begin{eqnarray}
  \label{eq:R}
  R(u) &=& 
  \left(\begin{array}{llll}
      \sin(\lambda +u) & 0 & 0 & 0 \\
      0 & \sin(u) & e^{-iu}\sin(\lambda)  & 0 \\
      0 & e^{iu}\sin(\lambda) & \sin(u) & 0 \\
      0 & 0 & 0 & \sin(\lambda +u)
    \end{array}\right) \;,\\[4mm]
  R^{\iota_1}(u) &=& 
  \left(\begin{array}{llll}
      \sin(\lambda +u) & 0 & 0 & 0 \\
      0 & -\sin(u) & e^{-iu}\sin(\lambda)  & 0 \\
      0 & e^{iu}\sin(\lambda) & -\sin(u) & 0 \\
      0 & 0 & 0 & \sin(\lambda +u)
    \end{array}\right).
\end{eqnarray}
(Notice that we introduced here the off-diagonal factors $e^{iu}$ and
$e^{-iu}$  not present in \cite{APSS} to allow the
decomposition (\ref{eq:deBaxterise}). They are nothing more than a rescaling
of the states or a simple gauge transformation.)
\\
A solution of
(\ref{eq:YBEconst1})--(\ref{eq:YBEconst4}) is then
given by 
\begin{eqnarray}
  \label{eq:constR}
  R &=& 
  \left(\begin{array}{llll}
      q & 0 & 0 & 0 \\
      0 & 1 & 0  & 0 \\
      0 & q-q^{-1} & 1 & 0 \\
      0 & 0 & 0 & q
    \end{array}\right) \;,\\[4mm]
  R^{\iota_1} &=& 
  \label{eq:constRi1}
  \left(\begin{array}{llll}
      q & 0 & 0 & 0 \\
      0 & -1 & 0  & 0 \\
      0 & q-q^{-1} & -1 & 0 \\
      0 & 0 & 0 & q
    \end{array}\right) \;,
\end{eqnarray}
where (\ref{eq:constR}) is the usual $R$-matrix of $\cU_q(gl(2))$.

\subsection{Algebra}

The YBE for $R$-matrices (\ref{eq:YBE1})--(\ref{eq:YBE2}) 
define the corresponding
YBE for $L$ and $L^{\iota}$-operators (the superscript
$\iota$ appeared according to definitions by the formula (\ref{T})), 
which act on quantum space of the chain.
According to formula (\ref{eq:deBaxterise}) one can introduce 
$L_a^{\pm}$,  ($a=1,2$) operators as
\begin{equation}
  \label{L}
  L_a(u)={1 \over 2i}(z L_a^+ - z^{-1} L_a^{-}).
\end{equation}
\medskip

\noindent
{\large \sc Proposition}. 
\textsl{The $R$-matrices (\ref{eq:constR}) 
and (\ref{eq:constRi1}) and the
equations (\ref{eq:YBEconst1})--(\ref{eq:YBEconst4}) lead to the
following algebra, defined by the generators $L^{\pm}$, 
$\left(L^{\pm}\right)^{\iota_1}$
\begin{eqnarray}
  \label{eq:RLL}
  R_{12} L_{1}^{\pm\iota_1} L_{2}^\pm &=&
  L_{2}^{\pm\iota_1} L_{1}^\pm R_{12}^{\iota_1} \\
  R_{12} L_{1}^{+\iota_1} L_{2}^- &=&
  L_{2}^{-\iota_1} L_{1}^+ R_{12}^{\iota_1} \\[3mm]
  R_{12}^{\iota_1} L_{1}^\pm L_{2}^{\pm\iota_1} &=&
  L_{2}^\pm L_{1}^{\pm\iota_1} R_{12} \\
  R_{12}^{\iota_1} L_{1}^+ L_{2}^{-\iota_1} &=&
  L_{2}^- L_{1}^{+\iota_1} R_{12} \\[3mm]
  R_{12} L_{1}^{-\iota_1} L_{2}^{+} 
  - \left(R_{21}\right)^{-1} L_{1}^{+\iota_1} L_{2}^{-} &=&
  L_{2}^{+\iota_1} L_{1}^{-} R_{12}^{\iota_1} 
  - L_{2}^{-\iota_1} L_{1}^{+} \left(R_{21}^{\iota_1}\right)^{-1} \\
  R_{12}^{\iota_1} L_{1}^{-} L_{2}^{+\iota_1} 
  - \left(R_{21}^{\iota_1}\right)^{-1} L_{1}^{+} L_{2}^{-\iota_1} &=&
  L_{2}^{+} L_{1}^{-\iota_1} R_{12} 
  - L_{2}^{-} L_{1}^{+\iota_1} \left(R_{21}\right)^{-1}. 
\end{eqnarray}
}

Writing the operators $L^\pm$  as usual in the form 
\begin{equation}
  \label{eq:Lmatrix}
  L^+ = \left(
    \begin{array}{cc}
      K_{+1} & 0 \\
      E & K_{+2} 
    \end{array}
  \right),
  \qquad
  L^- = \left(
    \begin{array}{cc}
      K_{-1} & F \\
      0 & K_{-2} 
    \end{array}
  \right)
\end{equation}
and similarly for $L^{\pm\iota_1}$, we get the relations

\begin{equation}
\label{K1}
\begin{array}{c}
 K_{+1}^{\iota_1} K_{-1} = K_{-1}^{\iota_1} K_{+1} 
 \\[2mm]
 K_{+2}^{\iota_1} K_{-2} = K_{-2}^{\iota_1} K_{+2} 
 \\[2mm]
 K_{+1}^{\iota_1} K_{+2} = - K_{+2}^{\iota_1} K_{+1} 
 \\[2mm]
 K_{-1}^{\iota_1} K_{-2} = - K_{-2}^{\iota_1} K_{-1} 
 \\[2mm]
 K_{+1}^{\iota_1} K_{-2} = - K_{-2}^{\iota_1} K_{+1} 
 \\[2mm]
 K_{+2}^{\iota_1} K_{-1} = - K_{-1}^{\iota_1} K_{+2} 
\end{array}
\qquad \qquad
\begin{array}{c}
 K_{+1} K_{-1}^{\iota_1}  = K_{-1} K_{+1}^{\iota_1} 
 \\[2mm]
 K_{+2} K_{-2}^{\iota_1}  = K_{-2} K_{+2}^{\iota_1}  
 \\[2mm]
 K_{+1} K_{+2}^{\iota_1}  = - K_{+2} K_{+1}^{\iota_1}  
 \\[2mm]
 K_{-1} K_{-2}^{\iota_1}  = - K_{-2} K_{-1}^{\iota_1}  
 \\[2mm]
 K_{+1} K_{-2}^{\iota_1}  = - K_{-2} K_{+1}^{\iota_1}  
 \\[2mm]
 K_{+2} K_{-1}^{\iota_1}  = - K_{-1} K_{+2}^{\iota_1}  
\end{array}
\end{equation}

\begin{equation}
\label{K2}
\begin{array}{c}
 K_{+1}^{\iota_1} E = q E^{\iota_1} K_{+1} 
 \\[2mm]
 K_{-1}^{\iota_1} E = q^{-1} E^{\iota_1} K_{-1} 
 \\[2mm]
 K_{+2}^{\iota_1} E = - q^{-1}  E^{\iota_1} K_{+2} 
 \\[2mm]
 K_{-2}^{\iota_1} E = - q E^{\iota_1} K_{-2} 
\end{array}
\qquad \qquad
\begin{array}{c}
 K_{+1} E^{\iota_1} = - q E K_{+1}^{\iota_1} 
 \\[2mm]
 K_{-1} E^{\iota_1} = - q^{-1} EK_{-1}^{\iota_1} 
 \\[2mm]
 K_{+2} E^{\iota_1} = q^{-1}  E K_{+2}^{\iota_1} 
 \\[2mm]
 K_{-2} E^{\iota_1} = q E K_{-2}^{\iota_1} 
\end{array}
\end{equation}

\begin{equation}
\label{K3}
\begin{array}{c}
 K_{+1}^{\iota_1} F = - q^{-1} F^{\iota_1} K_{+1} 
 \\[2mm]
 K_{-1}^{\iota_1} F = - q F^{\iota_1} K_{-1} 
 \\[2mm]
 K_{+2}^{\iota_1} F = q F^{\iota_1} K_{+2} 
 \\[2mm]
 K_{-2}^{\iota_1} F = q^{-1} F^{\iota_1}K_{-2} 
\end{array}
\qquad \qquad
\begin{array}{c}
 K_{+1} F^{\iota_1} = q^{-1} F K_{+1}^{\iota_1} 
 \\[2mm]
 K_{-1} F^{\iota_1} = q F K_{-1}^{\iota_1} 
 \\[2mm]
 K_{+2} F^{\iota_1} = - q F K_{+2}^{\iota_1} 
 \\[2mm]
 K_{-2} F^{\iota_1} = - q^{-1} FK_{-2}^{\iota_1} 
\end{array}
\end{equation}

\begin{eqnarray}
\label{K4}
  &&
  E^{\iota_1} F + F^{\iota_1} E = 
 (q-q^{-1}) \left( K_{-1}^{\iota_1} K_{+2} - K_{+1}^{\iota_1} K_{-2}
  \right) \;,
  \nonumber \\ && 
  E F^{\iota_1} + F E^{\iota_1} = - 
 (q-q^{-1}) \left( K_{-1} K_{+2}^{\iota_1} - K_{+1} K_{-2}^{\iota_1}
  \right) .
\end{eqnarray}

These algebraic relations look like those defining $\cU_q(gl(2))$, although
important sign differences appear, in particular in the exchange
relations of the generators $K$. We will refer to it below as
algebra~$\cA$.
Now we define the following quadratic operators
\begin{eqnarray}
  \begin{array}{rclrcl}
    e &=& K^{\iota_1}_1 E & \qquad\qquad
  f &=& K^{\iota_1}_2 F \\[1.5mm]
  k_1 &=& K^{\iota_1}_1 K_{1} &
  k_2 &=& K^{\iota_1}_2 K_{2} \\[1.5mm]
  l_1 &=& K^{\iota_1}_1 K_{-1} &
  l_2 &=& K^{\iota_1}_2 K_{-2} \\[1.5mm]
  m &=& K^{\iota_1}_1 K_{2}
  \end{array}
  \label{eq:composite}
\end{eqnarray}
It is easy to check that they satisfy the relations
\begin{eqnarray}
  \begin{array}{rclrcl}
    k_1 e &=& -q^{2} e k_1  &\qquad \qquad
    k_1 f &=& -q^{-2} f k_1  \\[1.5mm]
    k_2 e &=& -q^{-2} e k_2  &
    k_2 f &=& -q^{2} f k_2  \\[1.5mm]
    \left[e,f\right] &=& (q^2-1) (k_1 l_2 - k_2 l_1) \\[1.5mm]
    l_i e &=& - e l_i  &
    l_i f &=& - f l_i  \\[1.5mm]
    m e &=& - e m  &
    m f &=& - f m .
  \end{array}
  \label{eq:sl2qi}
\end{eqnarray}

The Cartan subalgebra of this deformed algebra is defined by the 
generators $k_1, l_1$ and $m$, while $k_2, l_2$ are being
fixed by central operators  $l_i^2$, $k_1 k_2 = -m^2$, $l_1l_2$, 
set to 1.
Therefore these relations correspond to the $\cU_{q,i}(gl(2))$ algebra, the
algebra with two deformation parameters $q$, $-1$. 
Usual definitions of multiparameter
deformations of the enveloping algebra $\cU(gl(2))$ were considered in
the articles \cite{RD}.

We can return back to the whole set of generators of the original
algebra (\ref{K1})--(\ref{K4}), inverting the  definitions
(\ref{eq:composite}) 
\begin{eqnarray}
  \begin{array}{rclrcl}
  E &=& K_{+1}    k_1^{-1}  e   & \qquad\qquad
  F &=& - K_{+1}    m^{-1}  f  \\[1.5mm]
  K_{-1} &=& K_{+1}    k_1^{-1}  l_1  &
  K_{+2} &=& K_{+1}    k_1^{-1}  m  \\[1.5mm]
  K_{-2} &=& - K_{+1}    m^{-1}  l_2  &
  E^{\iota_1} &=&  q^{-1} e    K^{-1}_{+1}   \\[1.5mm]
  F^{\iota_1} &=&  q k_1  m^{-1}  f    K^{-1}_{+1}   &
  K_{+1}^{\iota_1} &=& k_1    K^{-1}_{+1}   \\[1.5mm]
  K_{-1}^{\iota_1} &=& l_1    K^{-1}_{+1}   &
  K_{+2}^{\iota_1} &=& - m    K^{-1}_{+1}   \\[1.5mm]
  K_{-2}^{\iota_1} &=& k_1  m^{-1}  l_2   K^{-1}_{+1}. 
  \end{array}
\end{eqnarray}
For convenience, we define 
$E'= q E^{\iota_1}$,
$F'= - q^{-1} F^{\iota_1}$,
$K'_{\pm 1}= K_{\pm 1}^{\iota_1}$,
$K'_{\pm 2}= - K_{\pm 2}^{\iota_1}$
and write all the generators $A$ and $A'$ in the form
\begin{eqnarray}
  A &=& K_1 a \;, \nonumber\\
  A' &=& k_1 a K_1^{-1} \;.
\end{eqnarray}
Let $\cW$ be the associative algebra generated by $X$, $Z$, such that
$X^2=Z^2=1$ and $XZ+ZX=0$. ($\cW$ is equivalent to the algebra
satisfied by the Pauli matrices, which can be written $\sigma_1=X$,
$\sigma_3=Z$, and hence $\sigma^+=(1+Z)X/2$, $\sigma^-=(1-Z)X/2$. Then
\medskip
\\
\noindent
{\large \sc Theorem.} 
\textsl{The deformed algebra~$\cA$, defined by the relations
(\ref{K1})--(\ref{K4}) is isomorphic to  the tensor product
$\cW \otimes \cU_{q,i}(gl(2))$
}
\medskip

According to this theorem, the ordinary highest weight 
irreducible representations of our
algebra~$\cA$ are formed by direct product of $\CC^2$ 
with any irreducible representation 
of the deformed algebra $\cU_{q,i}(gl(2))$. One can draw the picture
as below expressing this fact.

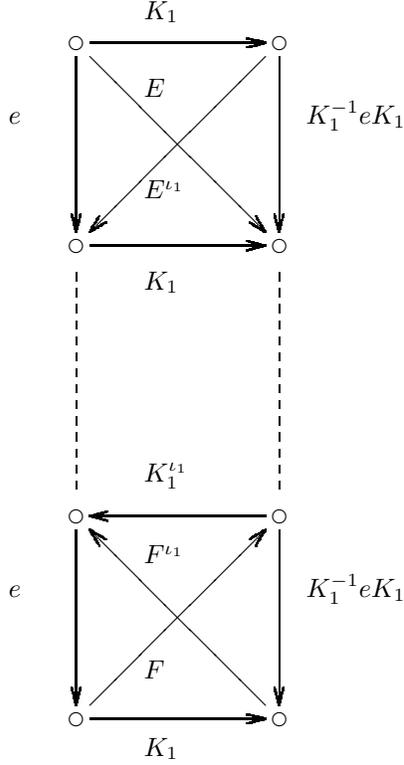
\setlength{\unitlength}{0.002 cm}

\begin{center}
\begingroup\makeatletter\ifx\SetFigFont\undefined%
\gdef\SetFigFont#1#2#3#4#5{%
  \reset@font\fontsize{#1}{#2pt}%
  \fontfamily{#3}\fontseries{#4}\fontshape{#5}%
  \selectfont}%
\fi\endgroup%
{\renewcommand{\dashlinestretch}{30}
\begin{picture}(2118,5055)(0,-10)
\put(450,4770){\circle{90}}
\put(1800,4770){\circle{90}}
\put(450,3420){\circle{90}}
\put(1800,3420){\circle{90}}
\put(450,1620){\circle{90}}
\put(1800,1620){\circle{90}}
\put(450,270){\circle{90}}
\put(1800,270){\circle{90}}
\drawline(540,4770)(1710,4770)
\drawline(1590.000,4740.000)(1710.000,4770.000)(1590.000,4800.000)
\drawline(540,4680)(1710,3510)
\drawline(1603.934,3573.640)(1710.000,3510.000)(1646.360,3616.066)
\drawline(1800,4680)(1800,3510)
\drawline(1770.000,3630.000)(1800.000,3510.000)(1830.000,3630.000)
\drawline(450,4680)(450,3510)
\drawline(420.000,3630.000)(450.000,3510.000)(480.000,3630.000)
\drawline(1710,4680)(540,3510)
\drawline(603.640,3616.066)(540.000,3510.000)(646.066,3573.640)
\drawline(540,3420)(1710,3420)
\drawline(1590.000,3390.000)(1710.000,3420.000)(1590.000,3450.000)
\dashline{60.000}(450,3240)(450,1800)
\dashline{60.000}(1800,3240)(1800,1800)
\drawline(450,1530)(450,360)
\drawline(420.000,480.000)(450.000,360.000)(480.000,480.000)
\drawline(540,270)(1710,270)
\drawline(1590.000,240.000)(1710.000,270.000)(1590.000,300.000)
\drawline(1800,1530)(1800,360)
\drawline(1770.000,480.000)(1800.000,360.000)(1830.000,480.000)
\drawline(646.066,1466.360)(540.000,1530.000)(603.640,1423.934)
\drawline(540,1530)(1710,360)
\drawline(1710,1620)(540,1620)
\drawline(660.000,1650.000)(540.000,1620.000)(660.000,1590.000)
\drawline(1646.360,1423.934)(1710.000,1530.000)(1603.934,1466.360)
\drawline(1710,1530)(540,360)
\put(900,4410){\makebox(0,0)[lb]{$E$}}
\put(900,3105){\makebox(0,0)[lb]{$K_1$}}
\put(900,4905){\makebox(0,0)[lb]{$K_1$}}
\put(900,0){\makebox(0,0)[lb]{$K_1$}}
\put(0,4230){\makebox(0,0)[lb]{$e$}}
\put(1980,4185){\makebox(0,0)[lb]{$K_1^{-1} e K_1$}}
\put(0,1080){\makebox(0,0)[lb]{$e$}}
\put(1980,1035){\makebox(0,0)[lb]{$K_1^{-1} e K_1$}}
\put(900,3735){\makebox(0,0)[lb]{$E^{\iota_1}$}}
\put(900,1800){\makebox(0,0)[lb]{$K_1^{\iota_1}$}}
\put(900,540){\makebox(0,0)[lb]{$F$}}
\put(900,1305){\makebox(0,0)[lb]{$F^{\iota_1}$}}
\end{picture}
}
\end{center}



We have here two columns of states forming irreps of $\cU_{q,i}(gl(2))$
with highest weight states $v_0$ and $v_1$. The operators marked by
the capital Latin 
letters ($K, E, F...$) maps from one column to the other, while small
letter operators ($e, f, k...$) act inside the columns.

It is known that in the case when $q$ is the root of unity the quantum
groups have a so called periodic (and semi-periodic) representations
(which are absent in the Lie algebras case). From the above construction 
of the representations, it follows that when
$q^r=1$ with $r\neq 4s, s=1,2...$, the periodic representations
here will not be a product of two irreps of $\cW$ and $\cU_{q,i}(gl(2))$,
but will form a joint irrep of double size of one for the
$\cU_{q,i}(gl(2))$ only.

\subsection{Alternative description}

Let us introduce now the direct sum of spaces $V_0$ and $V_1$
(formula (\ref{T})) as  $V=V_0 \oplus V_1$ and consider the following
operators acting there.
\medskip

\noindent
{\large \sc Definition.} 
\textsl{Let us define
\begin{equation}
  \label{eq:uK1}
  \cK_{\pm 1} = \left(
    \begin{array}{cc}
      0 & K_{\pm 1} \\
      K_{\pm 1}^{\iota_1} & 0
    \end{array}
  \right)
  \qquad
  \label{eq:uK2}
  \cK_{\pm 2} = \left(
    \begin{array}{cc}
      0 & K_{\pm 2} \\
      - K_{\pm 2}^{\iota_1} & 0
    \end{array}
  \right)
\end{equation}
\begin{equation}
  \label{eq:uE}
  \cE = \left(
    \begin{array}{cc}
      0 & E \\
      - E^{\iota_1} & 0
    \end{array}
  \right)
  \qquad
  \label{eq:uF}
  \cF = \left(
    \begin{array}{cc}
      0 & F \\
      F^{\iota_1} & 0
    \end{array}
  \right)
\end{equation}
\begin{equation}
  \label{eq:uL}
  \cB = \left(
    \begin{array}{cc}
      1 & 0 \\
      0 & -1
    \end{array}
  \right)
\qquad \qquad \cB^2 = 1
\end{equation}
}

After some simple matrix calculations we obtain the following
\medskip

\noindent
{\large \sc Proposition.} 
\textsl{The deformed algebra defined by the relations
(\ref{K1})--(\ref{K4}) can be represented as
\begin{eqnarray}
  \label{eq:commuL}
  \begin{array}{rclrcl}
    \cK_{\pm 1} \cK_{\pm 2} &=&  \cK_{\pm 2} \cK_{\pm 1} & \qquad\qquad
    \cB \cK_i &=& - \cK_i \cB \\[1.5mm]
    \cB \cE &=& - \cE \cB &
    \cB \cF &=& - \cF \cB \\[1.5mm]
    \cK_{\pm 1} \cE &=&  q^{\pm 1} \cB \cE \cK_{\pm 1} &
    \cK_{\pm 2} \cE &=&  q^{\mp 1} \cB \cE \cK_{\pm 2} \\[1.5mm]
    \cK_{\pm 1} \cF &=&  q^{\mp 1} \cB \cF \cK_{\pm 1} &
    \cK_{\pm 2} \cF &=&  q^{\pm 1} \cB \cF \cK_{\pm 2} \\[1.5mm]
    \left[\cE,\cF\right] &=& (q-q^{-1})\cB (\cK_2 \cK_{-1} - \cK_1 \cK_{-2})
  \end{array}
\end{eqnarray}
}

One can recognize in this relations the $\cU_q(gl(2))$
algebra, but instead of the ordinary deformation parameter $q$
as a complex number we have here a deformation matrix $q \cB$ with the property
$(q \cB)^2= q^2$. Therefore it is reasonable to mark algebra~$\cA$ 
as $sl_{q\cB}(2)$.

\section{Co-algebra structure}
\setcounter{equation}{0}

\indent

Because of the nature of the algebra $\cA$, defined by two
$R$-matrices and two sets of generators (with and without
$^{\iota_1}$), we cannot imagine a coproduct 
\begin{eqnarray}
  \label{eq:Delta2}
  \Delta &:& \cA\longrightarrow \cA\otimes \cA  \nonumber\\
  && L \longmapsto L \stackrel{.}{\otimes} L
\end{eqnarray}
as a morphism from $\cA$ to $\cA\otimes\cA$. Actually, the operators
obtained as $L \stackrel{.}{\otimes} L^{\iota_1}$ satisfy the commutation
relations of $\cU_q(gl(2))$, and we will use this fact later.
\\
We can however define two (related) notions: 

\subsection{Coproduct $\Delta^{(3)}$}
\medskip

\noindent
{\large \sc Definition.} \textsl{Define
\begin{eqnarray}
  \label{eq:Delta3}
  \Delta^{(3)} &:& \cA\longrightarrow \cA\otimes \cA \otimes \cA  \nonumber\\
  && L \longmapsto L \stackrel{.}{\otimes} L^{\iota_1}
  \stackrel{.}{\otimes} L 
\end{eqnarray}
i.e. 
\begin{eqnarray}
  \label{eq:Delta3gen}
  \Delta^{(3)}(K_1) &=& K_1 \otimes K_1^{\iota_1} \otimes K_1 \\
  \Delta^{(3)}(K_2) &=& K_2 \otimes K_2^{\iota_1} \otimes K_2 \\
  \Delta^{(3)}(E) &=& E \otimes K_1^{\iota_1} \otimes K_1
  + K_2 \otimes E^{\iota_1} \otimes K_1
  + K_2 \otimes K_2^{\iota_1} \otimes E \\
  \Delta^{(3)}(F) &=& K_{-1} \otimes K_{-1}^{\iota_1} \otimes F
  + K_{-1} \otimes F^{\iota_1} \otimes K_{-2}
  + F \otimes K_{-2}^{\iota_1}  \otimes K_{-2}
\end{eqnarray}
}
\medskip

\noindent
{\large \sc Proposition.} 
\textsl{$\Delta^{(3)}$ is a morphism of algebras. }

\subsection{Left co-action of $\cU_q(gl(2))$ on $\cA$}

Let us define
\begin{equation}
  \label{eq:uDelta}
  \underline\Delta : \cA \longrightarrow \cU_{q,i}(gl(2)) \otimes \cA  
\end{equation}
by 
\begin{eqnarray}
  \label{eq:uDeltagen}
  \underline\Delta(\cK_1) &=& k_1 \otimes \cK_1 \\
  \underline\Delta(\cK_2) &=& k_2 \otimes \cK_2 \\
  \underline\Delta(\cE) &=& e \otimes \cK_1 + k_2 \otimes \cE \\
  \underline\Delta(\cF) &=& {k}_1^{\iota_1} \otimes \cF + f \otimes 
\cK_2^{\iota_1} 
\end{eqnarray}
\medskip

\noindent
{\large \sc Proposition.} 
\textsl{$\underline\Delta$ is a morphism of algebras.}

\section{Generalization of $gl(N)$}
\setcounter{equation}{0}

\indent

Let us consider now the $gl(N)$ case.
The two constant $R$-matrices $R$ and $R^{\iota_1}$ given by 
\begin{equation}
  \label{eq:Rsln}
  R = \sum_{i=1}^{N} q e_{ii} \otimes e_{ii}
  + \sum_{i,j=1 \atop i\neq j}^{N}  e_{ii} \otimes e_{jj}
  + (q-q^{-1}) \sum_{i,j=1 \atop i>j}^{N} e_{ij} \otimes e_{ji}
\end{equation}
\begin{equation}
  \label{eq:Rslniota1}
  R^{\iota_1} = \sum_{i=1}^{N} q e_{ii} \otimes e_{ii}
  + \sum_{i,j=1 \atop i\neq j}^{N} b_{ij} e_{ii} \otimes e_{jj}
  + (q-q^{-1}) \sum_{i,j=1 \atop i>j}^{N} e_{ij} \otimes e_{ji}
\end{equation}
satisfy the four equations (\ref{eq:YBEconst1})--(\ref{eq:YBEconst4})
provided that  
\begin{equation}
  \label{eq:bij}
  b_{ij} = b_{ik} b_{kj} \qquad \mbox{and} \qquad b_{ij}^2=1
\end{equation}
This cocycle condition allows to write $b_{ij}=b_i b_j^{-1}$, with
$b_1=1$ and $b_i=\pm 1$ for $i>1$, thus yielding several
solutions. Let us now define the matrices 
\begin{equation}
  \label{eq:Bij}
  \cB_{ij} = \left(
    \begin{array}{cc}
      1 & 0 \\
      0 & b_{ij}
    \end{array}
  \right)
  = \cB_i \cB_j^{-1}\;,
  \qquad
  \cB_i = \left(
    \begin{array}{cc}
      1 & 0 \\
      0 & b_{i}
    \end{array}
  \right).
\end{equation}
\\
The $RLL$ relations (\ref{eq:RLL}) together with a gathering of the
operators as in (\ref{eq:Lmatrix}) 
lead to relations close to that of $\cU_q(gl(N))$ with some sign
modifications encoded in the $b_{ij}'s$. We denote by $\cA$ or 
$\cA_{q,\{b_j\}}$ the algebra generated by the $L$ and $L^{\iota_1}$ 
operators.
\\
We gather the operators $L^\pm$ and ${L^\pm}^{\iota_1}$ in matrices
\begin{equation}
  \label{eq:Lij}
  \cL^\pm_{ij} = \left(
    \begin{array}{cc}
      0 & {L^\pm}_{ij} \\
      b_i {{L^\pm}^{\iota_1}}_{ij} & 0
    \end{array}
  \right)
\end{equation}
which acts on a space $V= V_0 \oplus V_1$.
The equations 
$  R {L^\pm_1}^{\iota_1} L^\pm_2 =  L^\pm_2 {L^\pm_1}^{\iota_1} R^{\iota_1} $
and 
$  R^{\iota_1} {L^\pm_1} {L^\pm_2}^{\iota_1} =  
 {L^\pm_2}^{\iota_1} {L^\pm_1} R $
\ then read as
\begin{eqnarray}
  \label{eq:relationsslN}
  \begin{array}{rclc}
&&  \cB_{im}  \cL^\pm_{im} \cL^\pm_{in} 
    = q \cB_{in} \cL^\pm_{in} \cL^\pm_{im}   &  m<n 
    \\[2mm]
&&    \cB_{im} \cL^\pm_{im} \cL^\pm_{jm} = q \cB_{jm} \cL^\pm_{jm}
    \cL^\pm_{im}  &  i<j 
    \\[2mm]
&&    \cB_{im} \cL^\pm_{im} \cL^\pm_{jn} =  
    \cB_{jn} \cL^\pm_{jn} \cL^\pm_{im}  
    + (q-q^{-1})\; \cB_{jm} \cL^\pm_{jm} \cL^\pm_{in} & \quad i<j,\ m<n\\[2mm]
&&    \cB_{im} \cL^\pm_{im} \cL^\pm_{jn} =  
    \cB_{jn} \cL^\pm_{jn} \cL^\pm_{im}  
    &\quad  i<j,\ m>n \\[2mm]
&&    \cB_{im} \cL^\pm_{im} \cL^\pm_{jn}  =  
    \cB_{jn} \cL^\pm_{jn} \cL^\pm_{im}   & \quad i>j, \ m<n \\[2mm]
&&    \cB_{im} \cL^\pm_{im} \cL^\pm_{jn} 
    + (q-q^{-1})\; \cB_{jm} \cL^\pm_{jm} \cL^\pm_{in}  =  
    \cB_{jn} \cL^\pm_{jn} \cL^\pm_{im}  
    &\quad  i>j, \ m>n \\[2mm]
  \end{array}
\end{eqnarray}
while the equations 
$  R {L^+_1}^{\iota_1} L^-_2 =  L^-_2 {L^+_1}^{\iota_1} R^{\iota_1} $
and 
$  R^{\iota_1} {L^+_1} {L^-_2}^{\iota_1} =  {L^-_2}^{\iota_1} {L^+_1} R $
\ give
\begin{eqnarray}
  \label{eq:relationsslN2}
  \hspace{-1cm}
  \begin{array}{rclc}
&&    q \cB_{im} \cL^+_{im} \cL^-_{im}  =  
    q \cB_{im} \cL^-_{im} \cL^+_{im}  \\[2mm]
&&  \cB_{im}   \cL^+_{im} \cL^-_{in}  =  
    q^{-1} \cB_{in} \cL^-_{in} \cL^+_{im} + 
    q^{-1}(q-q^{-1})\; \cB_{im} \cL^-_{im}  \cL^+_{in} 
    & m<n \\[2mm]
&&  \cB_{im}   \cL^+_{im} \cL^-_{in}  =  
    q^{-1} \cB_{in} \cL^-_{in} \cL^+_{im}  & m>n \\[2mm]
&&  q^{-1} \cB_{im} \cL^+_{im} \cL^-_{jm} =  
    \cB_{jm} \cL^-_{jm} \cL^+_{im}  & i<j \\[2mm]
&&  q^{-1} \cB_{im} \cL^+_{im} \cL^-_{jm} + 
    q^{-1}(q-q^{-1})\; \cB_{jm} \cL^+_{jm} \cL^-_{im}  =  
    \cB_{jm} \cL^-_{jm} \cL^+_{im}  & i>j \\[2mm]
&&    \cB_{im} \cL^+_{im} \cL^-_{jn}  =  
    \cB_{jn} \cL^-_{jn} \cL^+_{im}  + (q-q^{-1})\; \cB_{jm} \cL^-_{jm}
    \cL^+_{in} 
    &  \quad i<j,\ m<n \\[2mm]
&&    \cB_{im} \cL^+_{im} \cL^-_{jn} =  
    \cB_{jn} \cL^-_{jn} \cL^+_{im} 
    &  \quad i<j,\ m>n \\[2mm]
&&    \cB_{im} \cL^+_{im} \cL^-_{jn} + 
    (q-q^{-1})\; \cB_{jm} \cL^+_{jm} \cL^-_{in}  =  
    \cB_{jn} \cL^-_{jn} \cL^+_{im}  
    + (q-q^{-1})\; \cB_{jm} \cL^-_{jm} \cL^+_{in}
    &  \quad i>j,\ m<n \\[2mm]
&&    \cB_{im} \cL^+_{im} \cL^-_{jn} + 
    (q-q^{-1})\; \cB_{jm} \cL^+_{jm} \cL^-_{in}  =  
    \cB_{jn} \cL^-_{jn} \cL^+_{im}  
    &  \quad i>j,\ m>n 
  \end{array}
  \nonumber \\
\end{eqnarray}


It is easy to see that if  $\cB_{ij}= 1$, i.e. if all $b_i$ are 1, 
then this 
set of algebraic relations simply becomes the set of definition
relations of the  quantum algebra $\cU_q(gl(N))$.
\\
In order to extract the $\cU_q(gl(N))$ part of this equations, let us
introduce the operator $\cM$, defined by the relation
\begin{equation}
  \label{MM}
\cM \cB_{ij} \cL_{ij} =\cL_{ij}\cB_{ij}.        
\end{equation}  

The simple expertise of the equations (\ref{eq:relationsslN})
and (\ref{eq:relationsslN2}) shows that the $\cB_{ij}$
matrices always appear there in the first position of 
the products and with the
same indices as the first operator $\cL_{ij}$. Therefore, by 
multiplying all equations by $\cM$ from the left and right hand
sides and using the relation (\ref{MM}) one can absorb the 
matrices $\cB_{ij}$ into operators 
$\bar{\cL}_{ij}= \cL_{ij} \cM$, for which we get the set
of $\cU_q(gl(N))$ defining relations.

The solution of the equations (\ref{MM}) can be written in the following
form. Let us denote as before $l_1=K_1^{\iota_1} K_{-1}=
L_{11}^{+\;\iota_1}L_{11}^{-}$, 
$k_1=K_1^{\iota_1} K_{1}=L_{11}^{+\;\iota_1}L_{11}^{+}$ and also 
$l_1^{\iota_1}=K_1 K_{-1}^{\iota_1}$, $k_1^{\iota_1}=K_1 K_{1}^{\iota_1}$.
We will consider in the following that the central operator $l_1^2$ is equal 
to 1.
One can check by direct calculations that the operator
\begin{eqnarray}
  \label{1MM}
  \cM = \left( \begin{array}{ll}
      0 & K_1 ({k_1}{l_1})^{-1/2}\nn\\
      ({k_1}{l_1})^{-1/2} K_1^{\iota_1}& 0
    \end{array}
  \right)
\end{eqnarray}
is fulfilling the equations (\ref{MM}) for all pairs $(i,j)$.

The operators $\cL_{ij} \cM$ have the form 
$\un \otimes \bar L_{ij}$, with $\bar L_{ij}$ a generator of standard 
$\cU_q(gl(N))$.
Therefore we have proved the following

\medskip
\noindent
{\large \sc Proposition.} 
\textsl{
The algebra $\cA_{q,\{b_j\}}$ generated by the relations
(\ref{eq:relationsslN})
and (\ref{eq:relationsslN2}) is equivalent to
$\left(\un \otimes \cU_q(gl(N))\right)$ extended by the operator
$\cM$, satisfying the relations (\ref{MM}) with the generators of 
$\left(\un \otimes \cU_q(gl(N))\right)$.
This supplementary operator can be regarded
as an additional Cartan generator, encoding both the deformation 
parameters $\{b_j\}$ and the doubling of the representation spaces. 
}
\\[3mm]
Since the operator $\cM$ does not commute with the rest $\cU_q(gl(N))$,
it seems that the algebra $\cA_{q,\{b_j\}}$  can not be represented as
a direct product. However one can introduce the operators
$I\equiv K_1 (k_1 l_1)^{-1/2}$ and
$I^{\iota_1}\equiv (k_1 l_1)^{-1/2} K_1^{\iota_1} l_1^{\iota_1}$
(these operators are always 
defined in finite dimensional representations) satisfying
$I^{\iota_1}I+II^{\iota_1}=1$ and note that
\begin{eqnarray}
  \label{2M}
  \cM=
  \left(\begin{array}{ll}
      0 & I \\
      I^{\iota_1} & 0  
    \end{array}
  \right) 
  \left(\begin{array}{ll}
      l_1^{\iota_1} & 0 \\
      0 & 1 
    \end{array}
  \right) 
  = 
  \left(\begin{array}{ll}
      1 & 0 \\
      0 &  l_1
    \end{array}
  \right)
  \left(\begin{array}{ll}
      0 & I \\
      I^{\iota_1} & 0  
    \end{array}
  \right) \;.
\end{eqnarray}

Let us now consider $\cU_{q,\{b_j\}}(gl(N))$, defined  as
$\cU_{q}(gl(N))$ extended by addition of the generator 
\begin{equation}
  \label{eq:l1}
  \bar l_1 \equiv \prod_i b_i^{h_{\omega_{i+1}}-h_{\omega_{i}}} \;.
\end{equation}
In this last expression, 
$h_{\omega_{i}}$ is the Cartan generator related with the
fundamental weight $\omega_{i}=\sum{(A^{-1})}_{ij} h_j$. 
The commutation properties of $\bar l_1$ with the generators of
$\cU_{q}(gl(N))$ are naturally deduced from the
expression (\ref{eq:l1}), taking the usual relations involving the
Cartan $h_i$'s (however not supposed to be themselves in $\cU_{q}(gl(N))$). 
The generator
$\bar l_1$ will be  responsible for the additional
deformation parameter $-1$ in $\cU_{q,\{b_j\}}(gl(N))$\footnote{This
  extension is different from the standard multiparametric deformation
  of $gl(N)$.}, the parameters
$b_i$ encoding  which directions of the Cartan generators are
concerned with this deformation. (Note that the existence of this new
generator may change the classification of representations at roots of
unity, the size of some periodic representations being for instance
enlarged). 

\medskip

\noindent
{\large \sc Proposition.} 
\textsl{
The algebra generated by ${L_{ij}^\pm}$ and ${L_{ij}^\pm}^{\iota_1}$
is equivalent to 
$\cW\otimes \cU_{q,\{b_j\}}(gl(N))$
with $\cW$ defined in section \ref{sect:gl2case}.
}
\\
The equivalence is indeed provided by the isomorphism 
\begin{eqnarray}
  \label{eq:eq:isomPhi}
  \phi(\sigma^+ \otimes \un) &=& I \\
  \phi(\sigma^- \otimes \un) &=& I^{\iota_1} \\
  \phi(\un \otimes \bL_{ij}) &=& L'_{ij }I^{\iota_1} + I^{\iota_1} L_{ij }\\
  \phi(\un \otimes \bar l_1) &=& l_1 + l_1^{\iota_1}
\end{eqnarray}
where $\bL_{ij}$ 
denote the generators of the standard $\cU_{q}(gl(N))$, that we extend 
with  $\bar l_1$ defined as above, 
and $L'_{ij}\equiv b_i L_{ij}^{\iota_1} l_1^{\iota_1}$.
We can check that 
$[\phi(\sigma^\pm \otimes \un),\phi(\un \otimes \bL_{ij})]=0$. 
\medskip
\\
At the end let us make the following remark. It is easy to find out
from the algebraic equations (\ref{eq:relationsslN})
and (\ref{eq:relationsslN2}) that  multiplying some of them by  
$\cB_{im}$ we can bring all of them to the form, where $\cB_{ij}$'s
appears only  coupled with $q$. Then, as in the $\cU_q(gl(2))$
case, one can talk about a quantum algebra with matrix valued
deformation parameters
$q \cB_{ij}, \; ((q \cB_{ij})^2 = q^2)$.

\section{Acknowledgments}
\setcounter{equation}{0}

\indent

The authors A.S. and T.S. acknowledge the LAPTH for hospitality,
where this work was carried out. T.S. acknowledge also INTAS
grant 99-1459 and A.S grant 00-390 for partial financial support.
  

\end{document}